\documentclass[slac_one]{revtex4}
\usepackage{graphicx}
\usepackage{fancyhdr}
\usepackage{subfigure}
\begin{document}

\title{
Status of MICE} 

%

\author{A. D. Bross}
\affiliation{Fermi National Accelerator Laboratory, Batavia, IL60510, USA}
\author{D. M. Kaplan}
\affiliation{Illinois Institute of Technology, Chicago, IL 60616, USA}
\begin{abstract}
Muon ionization cooling is the only practical method for preparing
high-brilliance beams needed for a neutrino factory or muon collider.
The muon ionization cooling experiment (MICE)  under development at the
Rutherford Appleton Laboratory comprises a dedicated beamline to
generate a range of input emittance and momentum, with time-of-flight and
Cherenkov detectors to ensure a pure muon beam. A first measurement of
emittance is performed in the upstream magnetic spectrometer with a
scintillating-fiber tracker. A cooling cell will then follow, alternating
energy loss in liquid hydrogen with RF acceleration. A second spectrometer
identical to the first  and a particle identification system
will measure the outgoing emittance. 
Plans for measurements of emittance
and cooling are described.
\end{abstract}

\maketitle

\thispagestyle{fancy}


\section{INTRODUCTION} 
Over 25 years have passed since the proposal of ionization cooling~\cite{cooling}. This key enabling technology for intense stored muon beams will be demonstrated experimentally for the first time in the Muon Ionization Cooling Experiment (MICE). Potential applications include: (1)~neutrino factories, a powerful tool for the study of neutrino oscillation and leptonic CP violation~\cite{Lindner}; and (2)~muon colliders, capable of high-luminosity, multi-TeV, lepton-antilepton collisions as well as  precision studies of $s$-channel-produced Higgs bosons~\cite{MC}. 

MICE is an experimental program to establish  the feasibility and performance of ionization cooling. The approach is to measure precisely the emittance of a 140 to 240\,MeV/$c$  muon beam  both before and after an ionization-cooling cell. The MICE collaboration includes over 100 accelerator and particle physicists and engineers from Belgium, Bulgaria, China, Italy, Japan, the Netherlands, Switzerland, the UK, and the US. The experiment is currently under construction at the Rutherford Appleton Laboratory (RAL) in the UK.

\section{MICE OVERVIEW AND TECHNICAL CHALLENGES}

\begin{figure}[h]
\vspace{-.1in}
\subfigure
{\includegraphics[width=.475\linewidth,bb=0 25 800 445,clip]{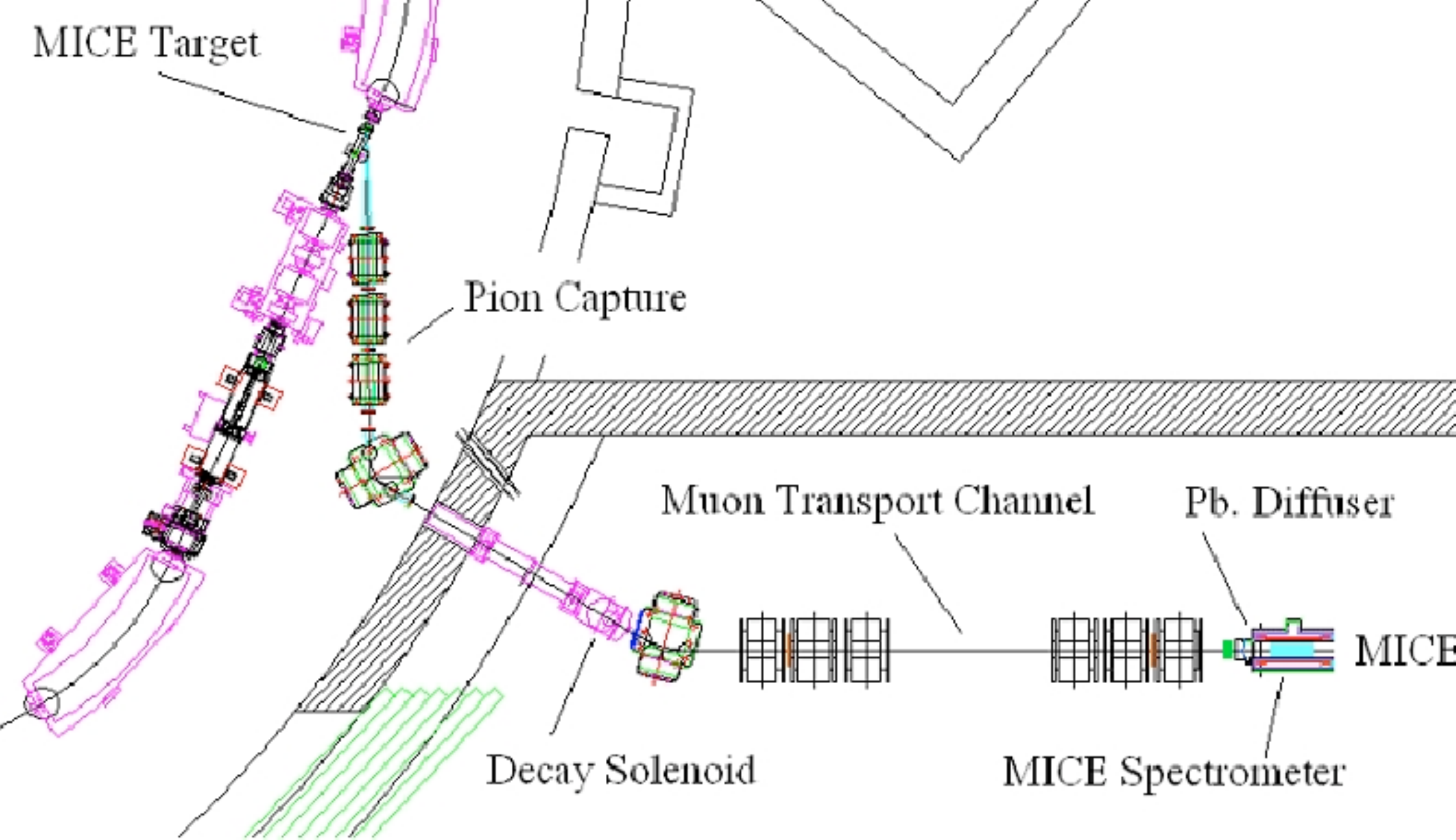}}
\subfigure
{\includegraphics[width=.475\linewidth,bb=22 85 749 452,clip]{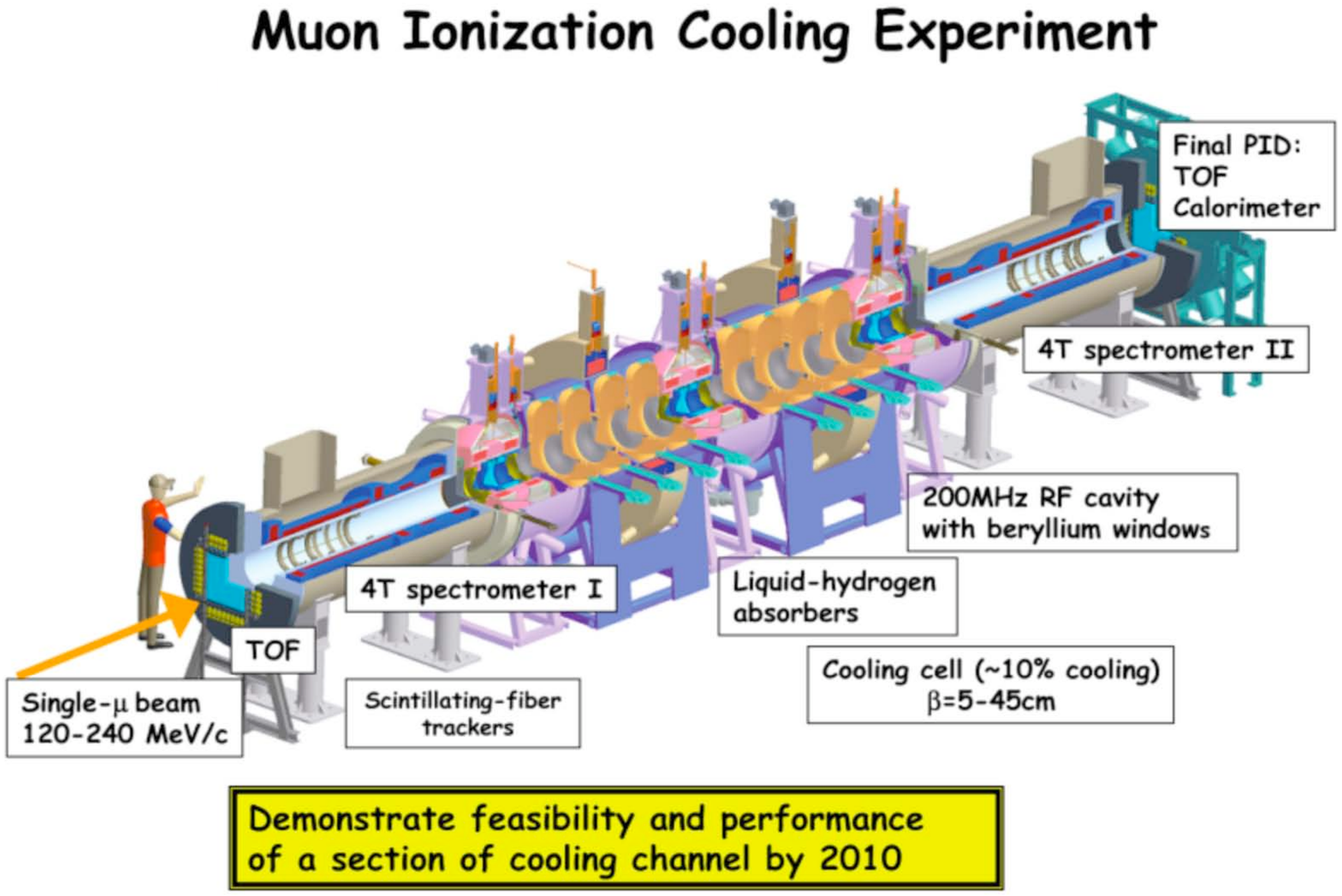}}
\vspace{-.1in}
\caption{(left) MICE beamline at ISIS, with 3 quadrupole triplets and two bends; (right) 3D rendering of MICE.}
\vspace{-.1in}
\label{fig:MICE}
\end{figure}

A muon beam produced by 800\,MeV protons from RAL's ISIS synchrotron is momentum-selected and transported to the MICE apparatus (see  Fig.~\ref{fig:MICE}). Particle identification ensures better than 99.9\% muon purity. The input beam is tunable (via magnet settings and a Pb diffuser of adjustable thickness) from 1 to 12$\pi$\,mm$\cdot$rad transverse emittance. The input 6D emittance is measured~\cite{Hart} in a magnetic spectrometer comprising a five-station scintillating-fiber tracker mounted within a 4\,T  superconducting (SC) solenoid. The tracker determines $x,x^\prime,y,y^\prime$, and particle energy, and time-of-flight (TOF) counters measure the sixth phase-space coordinate, $t$. 
The cooling cell consists of low-$Z$ absorbers and normal-conducting (NC) RF cavities, with SC coils providing strong focusing. The final emittance is measured in a second spectrometer system identical to the first one. Muon-decay electrons  would bias the emittance measurement and are eliminated via calorimetry.

Although conceptually simple, both ionization cooling and its experimental verification pose  challenges:
\begin{itemize}
\item Operation in strong  (1--3\,T)  magnetic fields of high-gradient ($\approx$16\,MV/m), low-frequency (201\,MHz) RF cavities. As shown by the Fermilab MuCool program, this exacerbates dark currents and cavity breakdown~\cite{MuCool}.
\item Safe designs employing substantial amounts of LH$_2$ near potential ignition sources (RF cavities).
\item The small effect ($\approx$10\%) of an affordable cooling device, which leads to the goal of $10^{-3}$ emittance precision, requiring single-particle measurements rather than standard beam instrumentation.
\end{itemize}

\section{CURRENT STATUS}

\begin{figure}[b]
\subfigure
{\includegraphics[width=.2\linewidth]{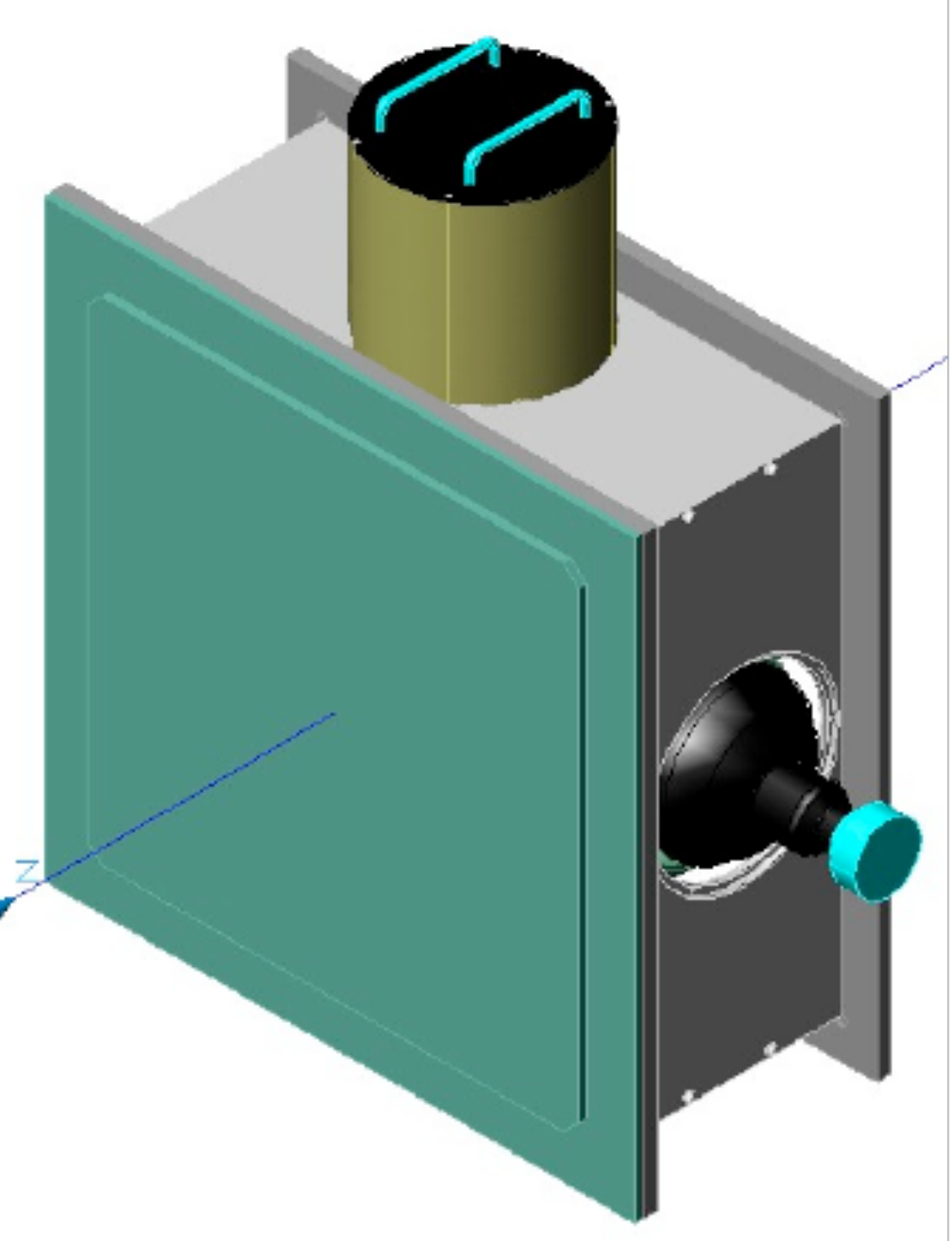}\includegraphics[width=.2\linewidth]{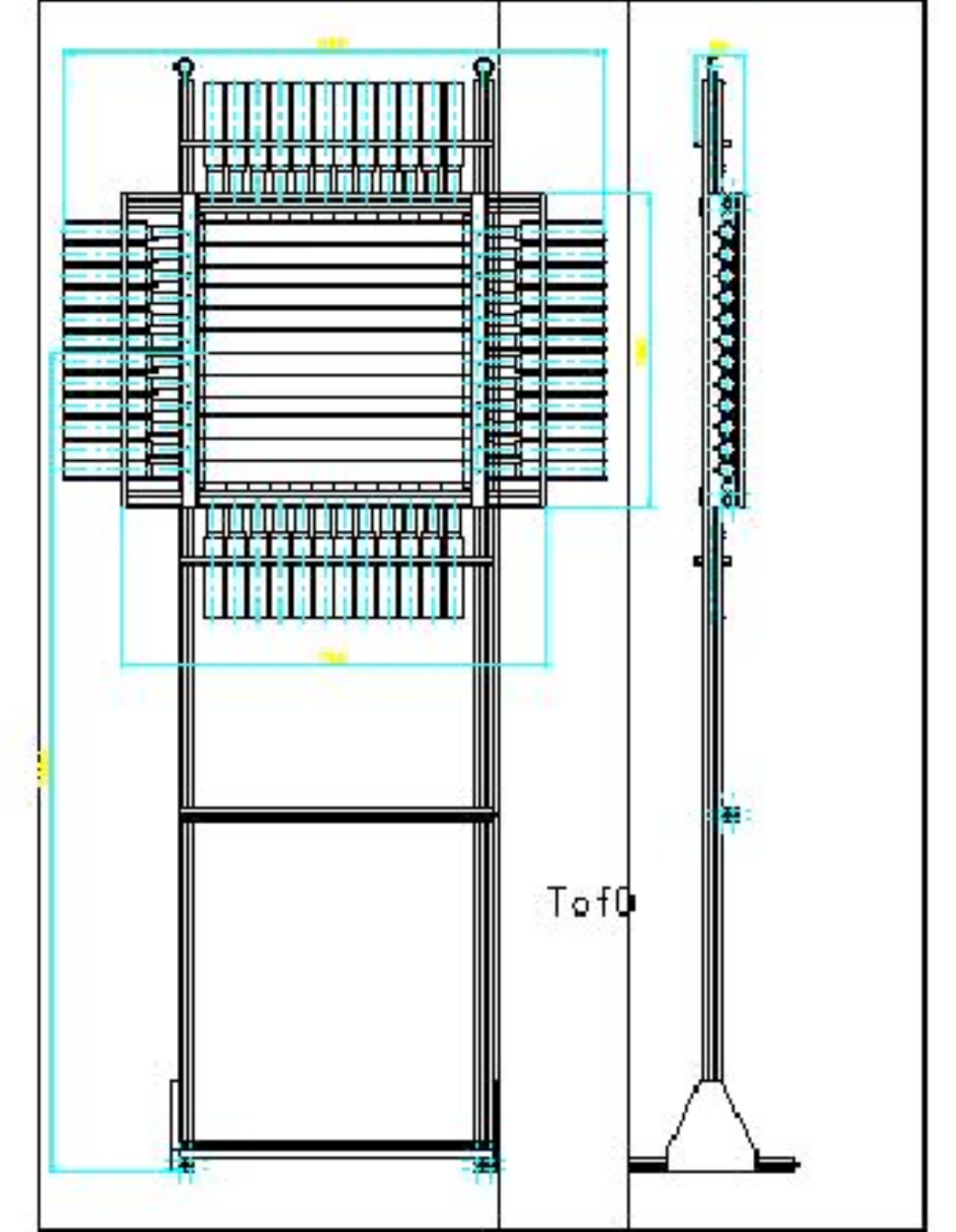}}
\subfigure
{\includegraphics[width=.55\linewidth]{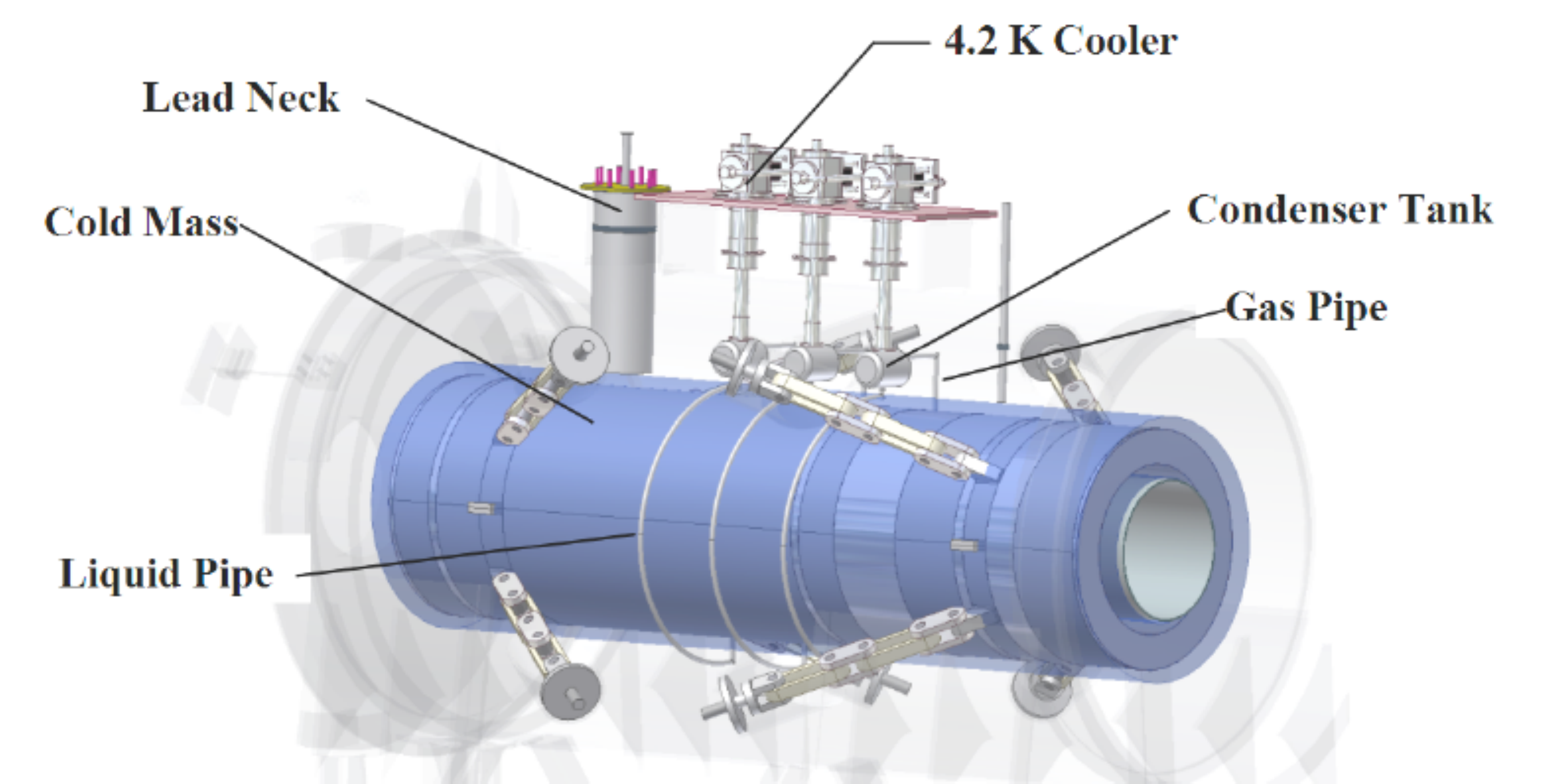}}
\vspace{-.1in}\caption{(left) MICE upstream (Cherenkov and TOF) particle-ID detectors and (right) schematic of spectrometer solenoid.}\label{fig:PID}
\end{figure}

The beamline (built by RAL; see Fig.~\ref{fig:MICE}) is now operational. Pions, from a small, movable Ti target (Sheffield Univ.)\ grazing the ISIS beam during its 2\,ms flat-top, are captured by a quadrupole triplet and momentum-selected by a dipole. Muons from decays within a 5-m-long, 12-cm-bore, 5\,T solenoid (donated by PSI, Switzerland; now in commissioning) are momentum-selected in a second dipole, at a momentum about half that of the first  (giving high muon purity). Three  two-layer scintillator-hodoscope TOF stations (Italy, assisted by Geneva and Sofia)  each provide 50\,ps resolution. The two (Belgium-US) aerogel Cherenkov counters and TOF0  (Fig.~\ref{fig:PID} left) are placed between quad triplets 2 and 3. Together, the TOF and Cherenkov counters distinguish $\pi$ from $\mu$ up to 300\,MeV/$c$. 

The 2-m-long spectrometer solenoids (US; Fig.~\ref{fig:PID} right) 
will provide 4\,T over a 1-m-long, 20-cm-radius tracking volume. Two ``end" coils ensure $<$1\% field nonuniformity; two ``matching" coils at one end match optics in and out of the cooling cell. The magnets are on order for 
late-2008 delivery, to be followed by measurement  and 
installation at RAL. Magnet sensors (NIKHEF) will monitor the field. Scintillating-fiber tracker 1 (Japan-UK-US) is complete and undergoing cosmic-ray testing; tracker 2 is in final assembly.  

Downstream particle ID comprises TOF2 and a calorimeter (Italy-Geneva-Sofia) to distinguish muons from decay electrons. A first, Pb--scintillating-fiber sandwich layer (similar to the KLOE calorimeter) precedes a  $\approx$1\,m$^3$, fully sensitive, segmented scintillator block. The sandwich layer degrades electrons; the scintillator block precisely measures muon range. Prototypes have been tested at Frascati and assembly of final detectors is in progress. 

\begin{figure}[t]
\vspace{-.25in}
\subfigure{
\includegraphics[width=.25\linewidth,bb=0 -15 400 500]{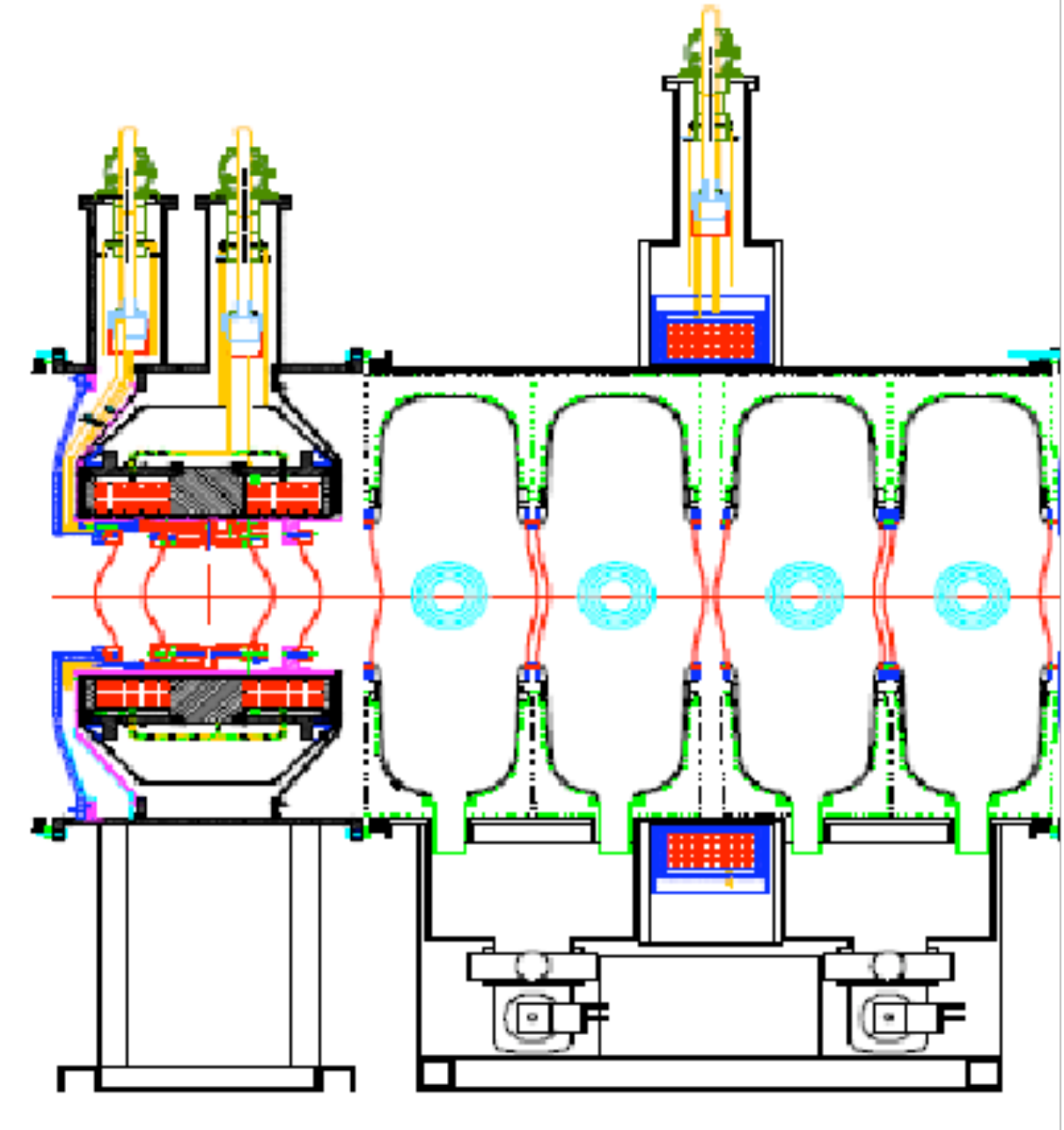}~~\includegraphics[width=.2\linewidth,bb=0 -25 330 500]{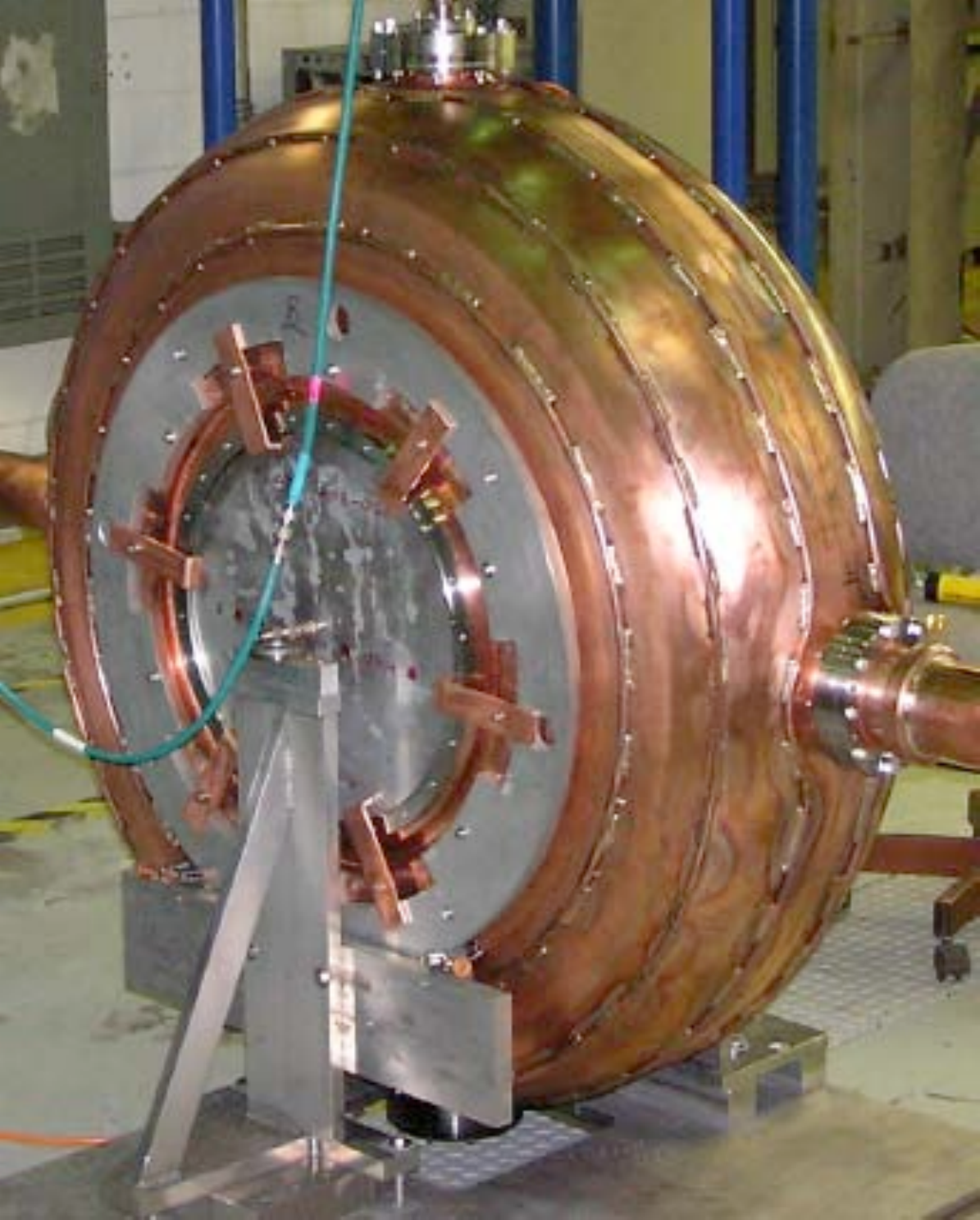}}\vspace{-.1in}
\subfigure{~
\includegraphics[width=.445\linewidth,bb=42 52 745 530,clip]{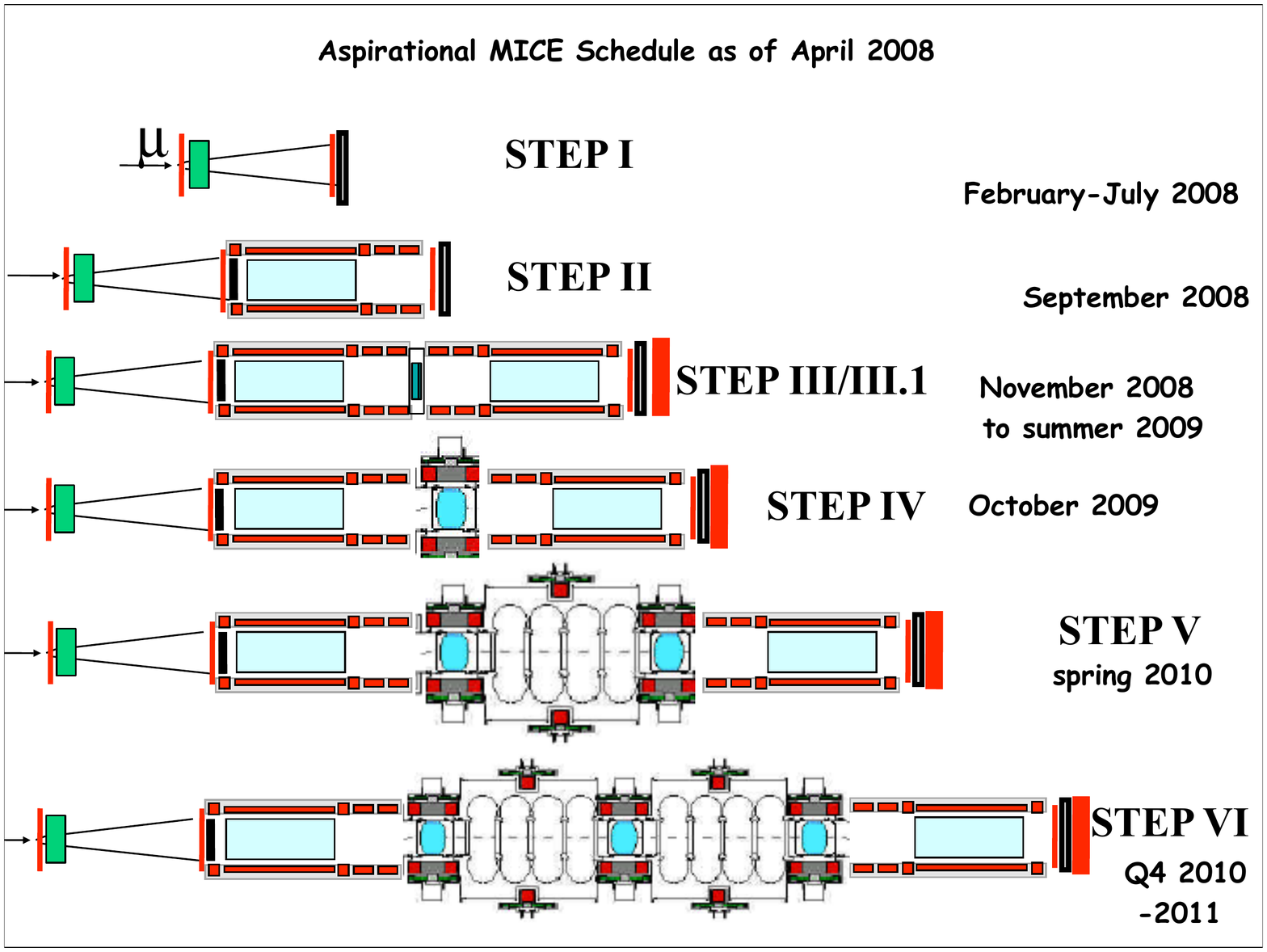}}
\vspace{-0.1in}
\caption{(left) Schematic of AFC module connected to RFCC module, with cavities closed by curved Be windows, and photo of prototype 201\,MHz cavity; (right) planned stages of MICE.}\label{fig:RF-sched}
\end{figure}

The absorber--focus-coil (AFC) modules (see Fig.~\ref{fig:RF-sched} left) will each
house an absorber, to reduce the muon momentum in all dimensions (and thereby cool the beam), and a pair of focusing coils, to reduce the beta function and hence the equilibrium emittance. While hydrogen is best, for emittance far from equilibrium, a more practical (though somewhat less efficient) absorption medium may be He or (solid) LiH or Be. MICE is therefore designed to test liquid and solid absorbers over a range of beta-function values.
 Critical  hydrogen-system issues  are (1) use of the thinnest possible metal containment windows, (2) safety, and (3) hydrogen storage.  A prototype  metal-hydride storage system  is in development at RAL. Safety reviews have been passed successfully, and the AFC modules are on order.

Since the  beam to be cooled is large, the RF frequency must be low; due to the focusing magnetic field, NC cavities are used. A large ``coupling" coil surrounds the cavities and, to minimize power requirements, Be windows  close them. Two RF--Coupling Coil (CC) modules connect three AFC modules  (Fig.~\ref{fig:RF-sched} left) to form the cooling cell. A CERN-refurbished 4\,MW RF-power source plus  two RF  sources donated by LBNL 
(in refurbishment at  Daresbury) give 8\,MW in total,  allowing  23\,MeV of acceleration in MICE Step VI. 
 Coupling Coil construction is commencing at the Institute of Cryogenics and Superconductivity Technology of the Harbin Institute of Technology (China), in collaboration with LBNL. Crucial high-magnetic-field cavity tests await delivery of the first CC. 

MICE will run at $\approx$1\,Hz and record  up to 600 muons in each 1\,ms beam burst. RF amplitude and phase, absorber mass, and other parameters must be precisely known for comparison with predictions and simulations. State-of-the-art instrumentation will monitor and record the relevant parameters (Daresbury-Geneva).

\section{SCHEDULE}
The staging of MICE (Fig.~\ref{fig:RF-sched} right) allows careful calibration at each step and also reflects funding realities.
Beam characterization and detector and data-acquisition run-in (Step I) have begun. 
Once the first spectrometer solenoid is available, a measurement of particle momenta (and hence emittance) will become possible (Step II). 
In Step III, up- and downstream emittance measurements will be precisely compared,  allowing a precise determination of  biases and testing of the required correction procedures. 
Step V will test ``repeatable" cooling, in which the momentum lost in the absorbers is restored in the RF cavities. Step VI will test the full cooling cell. 

\begin{acknowledgments}
The authors thank their collaborators of the MICE Collaboration. Work supported by Department of Energy under contract DE-AC02-76CH03000 and National Science Foundation grant PHY-031737.
\end{acknowledgments}

\end{document}